\documentclass[11pt,twoside]{article}
\usepackage{./asp2014}

\aspSuppressVolSlug
\resetcounters

\begin{document}

\title{Automated Extended Aperture Photometry for K2 RR Lyrae stars
}
\author{Emese Plachy,$^{1,2}$ P\'al Szab\'o,$^{1,2}$, Attila B\'odi,$^{1,2}$,  L\'aszl\'o Moln\'ar,$^{1,2}$ and R\'obert~Szab\'o$^{1,2}$}
\affil{$^1$ Konkoly Observatory, Research Centre for Astronomy and Earth Sciences, Konkoly Thege Mikl\'os \'ut 15-17, H-1121 Budapest
 }
\affil{$^2$ MTA CSFK Lend\"ulet Near-Field Cosmology Research Group;
\email{eplachy@konkoly.hu}}

\paperauthor{Emese Plachy}{eplachy@konkoly.hu}{0000-0002-5481-3352}{Research Centre for Astronomy and Earth Sciences}{Konkoly Observatory}{Budapest}{Budapest}{1121}{Hungary}
\paperauthor{P\'al Szab\'o}{szabopal96@gmail.com}{}{Research Centre for Astronomy and Earth Sciences}{Konkoly Observatory}{Budapest}{Budapest}{1121}{Hungary}
\paperauthor{Attila B\'odi}{bodi.attila@csfk.mta.hu}{0000-0002-8585-4544}{Research Centre for Astronomy and Earth Sciences}{Konkoly Observatory}{Budapest}{Budapest}{1121}{Hungary}
\paperauthor{L\'aszl\'o Moln\'ar}{molnar.laszlo@csfk.mta.hu}{0000-0002-8159-1599}{Research Centre for Astronomy and Earth Sciences}{Konkoly Observatory}{Budapest}{Budapest}{1121}{Hungary}
\paperauthor{R\'obert Szab\'o}{szabo.robert@csfk.mta.hu}{0000-0002-3258-1909}{Research Centre for Astronomy and Earth Sciences}{Konkoly Observatory}{Budapest}{Budapest}{1121}{Hungary}

\begin{abstract}  Light curves for RR Lyrae stars can be difficult to obtain properly in the K2 mission due to the similarities between the timescales of the observed physical phenomena and the instrumental signals appearing in the data. We developed a new photometric method called Extended Aperture Photometry (EAP), a key element of which is to extend the aperture to an optimal size to compensate for the motion of the telescope and to collect all available flux from the star before applying further corrections. We determined the extended apertures for individual stars by hand so far. Now we managed to automate the pipeline that we intend to use for the nearly four thousand RR Lyrae targets observed in the K2 mission. We present the outline of our pipeline and make some comparisons to other photometric solutions.
\end{abstract}

\section{The autoEAP pipeline}

The automated EAP pipeline is based on the EAP method \citep{eappaper} and consist of four basic steps. 
First we define initial stellar apertures with the \texttt{astropy} package.
Then we find the threshold to the number of times a given pixel is assigned to an aperture so that the highest number of stars are identified on the images.
The third step is to create light curves for all stars and identify the RR Lyrae variable from the Fourier parameters. Finally we generate a new set of apertures and select a new threshold which assigns the largest aperture to the RR Lyrae star whe the highest number of stars are detected in the image. The other apertures are then discarded. Steps may be repeated until only a single star is selected.

After we selected the aperture and generated the light curve, we apply the K2 Systematics Correction, or K2SC method \citep{k2sc1, k2sc2}, which can separate the pulsation from systematics effectively. 
 
\section{Comparison with other pipelines} 

We compared our results to other pipelines: the SAP and  PDCSAP light curves \citep{sap-pdcsap}, plus the K2SFF \citep{k2sff} and EVEREST \citep{everest} light curves. Three example stars with decreasing luminosities are presented in Fig.~1 (left to right). The last row shows the autoEAP solutions. In the high-luminosity case, the quality of the autoEAP light curve and EVEREST raw flux are comparable, though the former is slightly better. By visual inspection of the medium-luminosity star, autoEAP offers the best light curves, but on the other hand, in the low luminosity case, autoEAP quality is comparable to that of EVEREST again. It is our general observation that to maximize the likelihood of choosing a good-quality light curve, autoEAP offers the best choice. 

\section{Future plans} 
We are preparing the autoEAP light curves for the RR Lyrae stars in the K2 mission and will release the open-source autoEAP code as well. The code might be useful for not only RR Lyrae stars but other high-amplitude variable stars as well.

\articlefigure{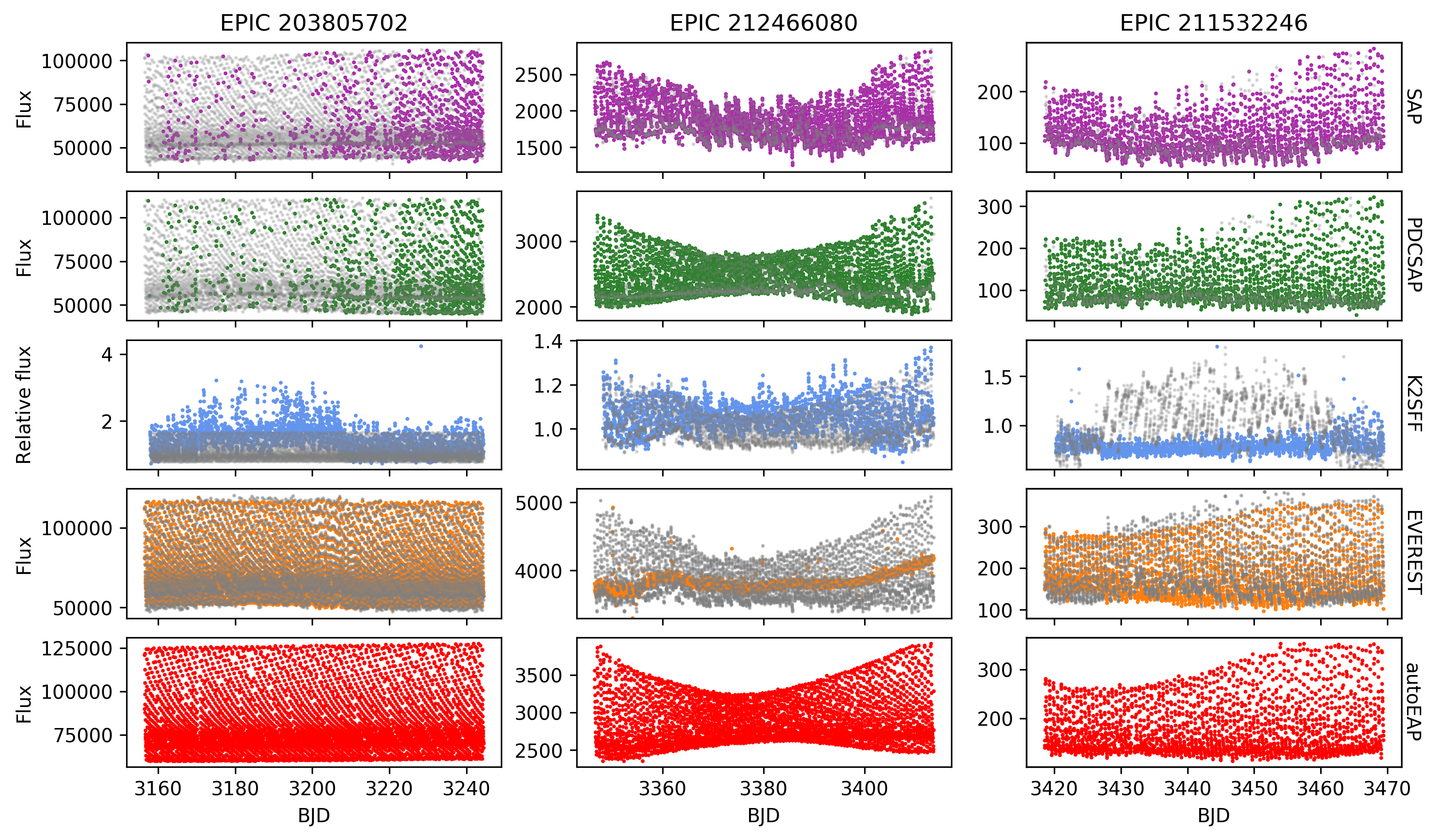}{ex_fig1}{Comparison of the pipelines. Gray: raw data; colored: corrected data.}

\acknowledgements Funding for the Kepler and K2 missions are provided by the NASA Science Mission directorate. EP was supported by the Bolyai J\'anos Research Scholarship, LM by the Premium Postdoctoral Research Program of the Hungarian Academy of Sciences. This research received funding from the NKFIH grants 2018-2.1.7-UK GYAK-2019-00009 and KH-18 (130405), and from the Lend\"ulet LP2018-7/2019 grants of the Hungarian Academy of Sciences. 



\end{document}